\def\half{\frac{1}{2}}

\def\bey{\begin{eqnarray}}
\def\eey{\end{eqnarray}}
\def\be{\begin{equation}}
\def\ee{\end{equation}}
\def\ba{\begin{array}}
\def\ea{\end{array}}
\def\gm{\gamma}

\def\sg{\sigma}
\def\Sg{\Sigma}

\def\om{\omega}
\def\r{\rho}

\def\dt{\delta}

\def\pp{\partial}
\def\pp{\partial}

\def\nnb{\nonumber}
\documentstyle[11pt,epsfig]{article}
\title{Equation of state of isospin-asymmetric nuclear matter
in relativistic mean-field models with chiral limits}
\author{Wei-Zhou Jiang$^{1,2}$, Bao-An Li$^{1}$, and Lie-Wen Chen$^{1,3}$  \\
\it  $^1$ Department of Physics, Texas A\&M University-Commerce,
Commerce, TX 75429, USA\\
\it $^2$ Institute of Applied Physics,
  Chinese Academy of Sciences, Shanghai 201800, China\\
\it $^3$ Institute of Theoretical Physics, Shanghai Jiao Tong
University, Shanghai 200240, China }
\date{}
\bigskip
\begin{document}
\maketitle \baselineskip 20.6pt

\begin{abstract}
\baselineskip 18pt Using in-medium hadron properties according to
the Brown-Rho scaling due to the chiral symmetry restoration at
high densities and considering naturalness of the coupling
constants, we have newly constructed several relativistic
mean-field Lagrangians with chiral limits. The model parameters
are adjusted such that the symmetric part of the resulting
equation of state at supra-normal densities is consistent with
that required by the collective flow data from high energy
heavy-ion reactions, while the resulting density dependence of the
symmetry energy at sub-saturation densities agrees with that
extracted from the recent isospin diffusion data from intermediate
energy heavy-ion reactions. The resulting equations of state have
the special feature of being soft at intermediate densities but
stiff at high densities naturally. With these constrained
equations of state, it is found that the radius of a 1.4$M_\odot$
canonical neutron star is in the range of 11.9 km$\leq$R$\leq$13.1
km, and the maximum neutron star mass is around 2.0$M_\odot$ close
to the recent observations.

\thanks Keywords: Equation of state, nuclear matter, symmetry energy,
relativistic mean-field models, chiral limits. PACS numbers:
21.65.+f, 26.60.+c, 11.30.Rd
\end{abstract}

\section{Introduction}
The Equation of State (EOS) of isospin asymmetric nuclear matter
plays a crucial role in many important issues in astrophysics, see,
e.g., Refs.~\cite{lat01,steiner05a,ho01}. It is also important for
understanding both the structure of exotic nuclei and the reaction
dynamics of heavy-ion collisions, see, e.g., Ref.~\cite{ba01}.
Within the parabolic approximation, the energy per nucleon in
isospin asymmetric nuclear matter can be written as
$E/A=e(\r)+E_{sym}(\r)\dt^2 $ where $e(\r)$ is the EOS of symmetric
nuclear matter, the $E_{sym}(\r)$ is the symmetry energy and
$\dt=(\r_n-\r_p)/\r$ is the isospin asymmetry. Both the $e(\r)$ and
the $E_{sym}(\r)$ are important in astrophysics although maybe for
different issues. For instance, the maximum mass of neutron stars is
mainly determined by the EOS of symmetric nuclear matter $e(\r)$
while the radii and cooling mechanisms of neutron stars are
determined instead mainly by the symmetry energy
$E_{sym}$\cite{lat01,li06}. The nuclear physics community has been
trying to constrain the EOS of symmetric nuclear matter using
terrestrial nuclear experiments for more than three decades, see,
e.g.\cite{da02}, for a review. On the other hand, a similarly
systematic and sophisticated study on the density dependence of the
symmetry energy $E_{sym}$ using heavy-ion reactions only started
about ten years ago stimulated mostly by the progress and
availability of radioactive beams\cite{lkr}. Compared to our current
knowledge about the EOS of symmetric nuclear matter, the symmetry
energy $E_{sym}$ is still poorly known especially at supra-normal
densities~\cite{li02} given the recent progress in constraining it
at densities less than about 1.2$\r_0$ using the isospin diffusion
data from heavy-ion reactions~\cite{ts04,ch05,li05}.

The aim of this work is to investigate the EOS of isospin asymmetric
nuclear matter within the relativistic mean-field (RMF) model with
in-medium hadron properties governed by the BR scaling. From the
point of view of hadronic field theories, the symmetry energy is
governed by the isovector meson exchange. Studying in-medium
properties of isovector mesons is thus of critical importance for
understanding the density dependence of the symmetry energy. We
first construct model Lagrangians respecting the chiral symmetry
restoration at high densities. The model parameters are adjusted
such that the symmetric part of the resulting EOS at supra-normal
densities is consistent with that required by the collective flow
data from high energy heavy-ion reactions~\cite{da02}, while the
resulting density dependence of the symmetry energy at
sub-saturation densities agrees with that extracted from the recent
isospin diffusion data from intermediate energy heavy-ion
reactions~\cite{ts04,ch05,li05}. The constrained EOS is then used to
investigate several global properties of neutron stars.

\section{Relativistic mean-field models with chiral limits}
In-medium properties of the isovector meson $\r$ can be studied
through the special symmetry breaking and restoration. The local
isospin symmetry in the Yang-Mills field theory, where the $\r$
meson may be introduced as a gauge boson of the strong interaction,
can serve as a possible candidate to study the in-medium properties
of $\r$ meson. However, since $\pi N$ interactions actually dominate
the strong interaction in hadron phase, it was rather difficult to
understand how the in-medium properties of the massive $\r$ meson
could be consistent with the restoration of local isospin symmetry.
On the other hand, within the microscopic theory for the strong
interaction, namely, the QCD which is a color SU(3) gauge theory,
the chiral symmetry is approximately conserved. The spontaneous
chiral symmetry breaking and its restoration can be manifested in
effective QCD models. Based on the latter, Brown and Rho (BR)
proposed the in-medium scaling law~\cite{br91} implying that hadron
masses and meson coupling constants in the Welacka model~\cite{se86}
approach zero at the chiral limit. The scaling was treated in the
hadronic phase before the chiral symmetry restoration.

As an effective QCD field theory, the hidden local symmetry theory
has been developed to include the $\r$ meson in addition to the pion
in the framework of the chiral perturbative theory by Harada and
Yamawaki~\cite{ha03,ha07} and it is shown that  the $\r$ meson
becomes massless at the chiral limit. This supports the mass
dropping senario of the BR scaling. There are also experimental
indication for the mass dropping, i.e., the dielectron mass spectra
observed at the CERN SPS~\cite{ag95,li95}, the $\om$ meson mass
shift measured at the KEK~\cite{kek} and the ELSA-Bonn~\cite{tap},
as well as the analysis of the STAR data~\cite{sh03,br07}. However,
data from the NA60 Collaboration for the dimuon spectrum~\cite{na60}
seem to favor the explanation of $\r$ meson broadening based on a
many-body approach~\cite{ra06}. So far, the controversy is still
unsettled~\cite{br07}. The chiral symmetry and its spontaneous
breaking are closely related to the mass acquisition and dropping of
hadrons. Since the chiral symmetry is a characteristic of the strong
interaction within the QCD, it is favorable to include in the RMF
models effects of the chiral symmetry through the BR scaling law.
However, this does not mean that the contribution of the many-body
correlations~\cite{hof01} is excluded. Actually, the contribution of
the many-body correlations can be included phenomenologically into
the RMF models to reproduce the saturation properties of nuclear
matter.

The in-medium $\r$ meson plays an important role in modifying the
density dependence of the symmetry energy. For most RMF models
that the $\r$ meson mass is not modified by the medium, the
symmetry energy is almost linear in  density.  The introduction of
the isoscalar-isovector coupling in RMF models  can    soften  the
symmetry energy at high densities~\cite{ho01}.  Meanwhile, it
reproduces the neutron-skin thickness in $^{208}$Pb as that given
by the non-relativistic models (about 0.22fm)~\cite{fsu},
consistent with the available data~\cite{kl07}. In well-fitted RMF
models that give a value of incompressibility $\kappa=230$ MeV, a
large coefficient of non-linear self-interacting $\om$ meson term
is required~\cite{fsu} and thus the naturalness breaks down.
Moreover, the isoscalar-isovector coupling in the RMF models
increases the effective $\r$ meson mass with density, which leads
the model to be far away from the chiral limit.

The Walecka model with the density-dependent parameters is the
simplest version to incorporate the effects of chiral symmetry. The
Lagrangian is written as \bey
 {\mathcal{ L}}&=&
{\overline\psi}[i\gm_{\mu}\partial^{\mu}-M^* +g^*_{\sg}\sg-g^*_{\om }
\gm_{\mu}\om^{\mu}-g^*_{\r}\gm_\mu \tau_3 b_0^\mu]\psi
 +\frac{1}{2}(\partial_{\mu}\sg\partial^{\mu}\sg-m_{\sg}^{*2}\sg^{2})
 \nnb\\
 && - \frac{1}{4}F_{\mu\nu}F^{\mu\nu}+
      \frac{1}{2}m_{\om}^{*2}\om_{\mu}\om^{\mu}
       - \frac{1}{4}B_{\mu\nu} B^{\mu\nu}+
\frac{1}{2}m_{\r}^{*2} b_{0\mu} b_0^{\mu} \label{eq:lag1}   \eey
where $\psi,\sg,\om$, and  $b_0$  are the fields of the nucleon,
scalar, vector, and  isovector-vector mesons, with their in-medium
scaled masses $M^*, m^*_\sg,m^*_\om$, and $m^*_\r$, respectively. The
meson coupling constants and  hadron masses with asterisks denote the
density dependence, given by the BR scaling.  The energy density and
pressure read, respectively,
 \bey
 {\mathcal{E}}&=&\half C_\om^2\r^2+\half C_\r^2 \r^2\dt^2+ \half
 \tilde{C}^2_\sg(m_N^*-M^*)^2 +\sum_{i=p,n}
 \frac{2}{(2\pi)^3}\int_{0}^{{k_F}_i}\! d^3\!k~ E^*,\\
  \label{eqe1}
 p&=&\half C_\om^2\r^2+\half C_\r^2 \r^2\dt^2- \half
 \tilde{C}^2_\sg(m_N^*-M^*)^2 -\Sg_0\r+
 \frac{1}{3}\sum_{i=p,n}\frac{2}{(2\pi)^3}\int_{0}^{{k_F}_i}\! d^3\!k
 ~\frac{{\bf k}^2}{E^*}
\label{eqp1}
 \eey
where $C_\om=g_\om^*/m_\om^*$, $C_\r=g_\r^*/m_\r^*$,
$\tilde{C}_\sg=m_\sg^*/g_\sg^*$,   $E^*=\sqrt{{\bf k}^2+{m_N^*}^2}$
with $m_N^*=M^*-g^*_\sg\sg$ the effective mass of nucleon, and $k_F$
is the Fermi momentum. The incompressibility of symmetric matter can
be expressed explicitly as~\cite{iw00}
\begin{equation}\label{kappa}
\kappa=9\r\frac{\pp\mu}{\pp\r}=9\r(C^2_\om+2C_\om\r \frac{\pp
C_\om}{\pp\r} +\frac{\pp E_F}{\pp\r}-\frac{\pp\Sg_0}{\pp\r})
\end{equation}
where the chemical potential is given by $\mu=\pp{\mathcal E}/\pp\r$
and the Fermi energy is $E_F=\sqrt{k_F^2+{m_N^*}^2}$. The
rearrangement term is essential for the thermodynamic consistency to
derive the pressure in (\ref{eqp1}) and its expectation value in the
mean field $\Sg_0$ is given by
\begin{equation}\label{rre}
    \Sg_0=-\r^2C_\om \frac{\pp C_\om}{\pp \r}-\r^2\dt^2 C_\r\frac{\pp C_\r}{\pp
    \r}-\tilde{C}_\sg\frac{\pp\tilde{C}_\sg}{\pp \r}(m_N^*-M^*)^2-
    \r_s\frac{\pp M^*}{\pp \r}.
\end{equation}
The density dependence of parameters is usually described by the
scaling functions  that are the ratios of the in-medium parameters to
those in the free space. The choice of scaling functions and their
coefficients are constrained by the saturation properties of nuclear
matter and experimental data about the in-medium mass dropping of
vector mesons. Moreover, we also use as a constraint the pressure
within the density range $2-4.6\r_0$ extracted from measurements of
nuclear collective flows in heavy-ion collisions~\cite{da02}. The
scaling function may take the form~\cite{song01}: \be
\Phi(\r)=\frac{1}{1+y\r/\r_0}\label{sc1}
 \ee
with $y=0.28$ for the vector meson mass, giving $\Phi(\r_0)=0.78$
found in QCD sum rules~\cite{jin96}. Recently, in Ref.~\cite{sc07}
where the memory effect in dimuon yield was studied by considering
the mass dropping of $\r$ meson, the authors cited a scaling
function~\cite{br05}
 \be
 \Phi(\r)=1-y\r/\r_0\label{sc2}
 \ee
with $y=0.15$. Data extracted from the $\gamma-A$ reaction by the
TAP collaboration indicate a value of $y\approx 0.13$~\cite{tr05}.
In addition, data from the KEK photon-induced nuclear reaction
indicate that the $\om$ meson mass dropping is about the same
order of magnitude at the saturation density~\cite{na06}.

Song firstly built effective models based on the BR scaling using the
scaling functions (\ref{sc1}) for both hadron masses and vector
coupling constants~\cite{song01}. A reasonable incompressibility for
nuclear matter was obtained by introducing the non-linear
self-interacting meson terms with coefficients satisfying the
hypothesis of {\it naturalness} that is originated from the chiral
symmetry and QCD scaling~\cite{ma84,ma04,ue06,we90}. In
Ref.~\cite{liu04}, along the line of \cite{song01}, the BR scaling
function (\ref{sc1}) was considered only for hadron masses with a
small value of $y$ in the RMF models to study nuclear matter
properties. In a more recent work~\cite{av06}, the BR scaling
function (\ref{sc1}) was taken for the scalar and vector meson
coupling constants with respective values of $y$, and the scaling
function (\ref{sc2}) was taken for hadron masses in the RMF models
without the self-interacting meson interactions. Though the models
built in these works can give rather good descriptions of nuclear
saturation properties, the pressures calculated in the density region
of $\r=2-4.6\r_0$, however, are still far away from that extracted
from measurements of nuclear collective flows in heavy-ion
collisions~\cite{da02}. People have already tried to improve the
situation by including the non-linear meson self-interacting terms.
Unfortunately, it is not satisfactory because even unreasonably large
coefficients of non-linear meson terms that break down the hypothesis
of {\it naturalness} can not reduce the pressure lower enough to pass
the experimental pressure-density region.

The effective nucleon mass is not dominated by the BR scaling (at
least at the normal nuclear matter density) since it is usually
around $0.65M^*$ at normal density while the mass dropping given
by the BR scaling is less than 15\% ($y\leq0.15$ in (\ref{sc2})).
This implies that the scalar meson coupling constant that plays a
crucial role in the effective nucleon mass can adopt a different
density-dependent scaling from that of the vector meson. In
particular, the high pressure predicted by various models at high
densities should be lowered, and this requires the scalar coupling
constant to decrease more slowly. Furthermore, if the scaling
function (\ref{sc2}) for the $\om$ meson mass is preferred by
experiments, one may take the same scaling function (\ref{sc2})
for the coupling constant of $\om$ meson  to avoid the infinity of
pressure at the chiral limit where the scalar density vanishes. In
this way, the effect of density dependence from the vector meson
part is cancelled out since the energy density (\ref{eqe1}) only
relies on the ratios $C_\om$ and $C_\r$. Generally, we may take
the scaling functions for the coupling constants of vector mesons
as
\begin{equation}\label{sc3}
   \Phi_\r(\r)=\frac{1-y\r/\r_0}{1+y_\r\r/\r_0}, \hbox{ }
   \Phi_\om(\r)=\frac{1-y\r/\r_0}{1+y_\om\r/\r_0}.
\end{equation}
For  hadron masses (including nucleons, if they have),  (\ref{sc2})
is taken with the same value of $y$ used in (\ref{sc3}). For the
$\sg$ meson coupling constant,   the same form as (\ref{sc1}) is
taken but with a coefficient denoted by $x$:
\begin{equation}\label{sc4}
     \Phi_\sg(\r)=\frac{1}{1+x\r/\r_0}.
\end{equation}

\section{Results and discussions}
\begin{table}
\caption{Parameter sets fitted at the saturation density
$\r_0=0.16$fm$^{-3}$. The vacuum hadron masses are $M=938$MeV,
$m_\sg=500$MeV, $m_\om=783$MeV and $m_\r=770$MeV except for
$m_\sg=600$MeV for the parameter set S3. The coupling constants given
here are those at zero density. For parameter sets SL3 and S3, the
non-linear $\sg$ self-interacting coefficients are introduced (see
text). The parameter set SL1$^*$ has two more parameters $y_\r=0.654$
and $y_\om=0.0365$. The symmetry energy is fitted to 31.6MeV at
$\r=0.16$fm$^{-3}$ for all models. The critical density $\r_c$ for
the chiral symmetry restoration is given by  the value $y$ in
(\ref{sc2}) for zero hadron mass. \label{t:t1}}
 \begin{center}
    \begin{tabular}{ c c c c c c c c c c c c c}
\hline\hline &$g_\sg$ & $g_\om$ & $g_\r$ & $y$ & $x$  &
$\kappa$(MeV)   & $M^*/M$ & $M^*_n/M$& $\r_c/\r_0$    \\
\hline
SL1  &8.6388 &10.4634 &3.7875 &0.126 & 0.234 &230.0 & 1.0& 0.679&7.94   \\
SL1$^*$  &9.7414 &12.5535 &5.8644 &0.126 &0.238 &230.0 & 1.0& 0.600&7.94\\
SL2  &6.1664 &10.9682 &3.9866 &0.11 & 0.381 &219.5 &0.89 &0.763&9.09  \\
SL3  &9.8627 &12.4928 &3.6128 &0.126 & - &250.0 &1.0 &0.620 &7.94 \\
S3\cite{song01} &5.3210 &15.3134 & 3.6035 &0.28 & - &250.0 &0.78 &0.617 & - \\
\hline\hline
\end{tabular}
\end{center}
\end{table}

\begin{figure}[tbh]
\begin{center}
\includegraphics[width=0.7\textwidth]{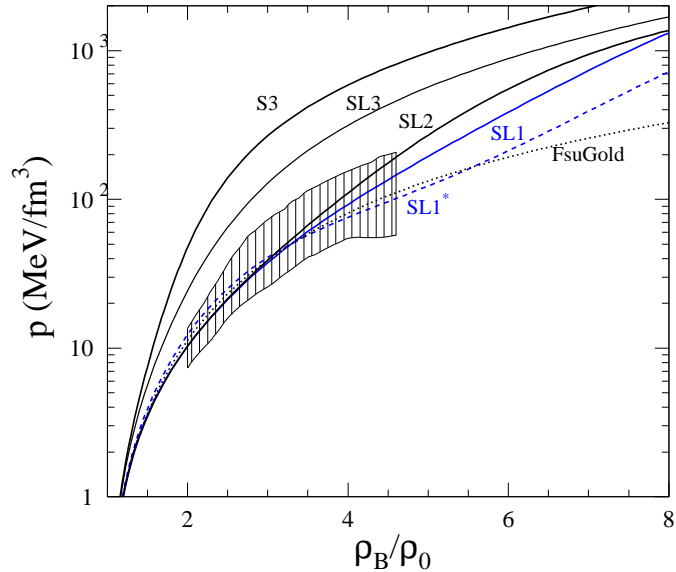}
 \end{center}
\caption{The pressure as a function of density for different models.
The shaded region is given by experimental error bars\cite{da02}.
\label{pden}}
\end{figure}

We first adjust the parameters to reproduce the saturation properties
including the binding energy per nucleon $E/A-M=-16$ MeV, the zero
pressure, the incompressibility $\kappa$ and the effective nucleon
mass $m_N^*$ at  saturation density $\r_0=0.16$fm$^{-3}$. The
resulting parameter sets SL1 and SL2 and the corresponding saturation
properties are tabulated in Table \ref{t:t1}. In SL1 the nucleon mass
scaling is not considered and the effective nucleon mass is just
$m_N^*=M-g_\sg^*\sg$. In SL2, the nucleon mass scaling is included,
and a much larger effective nucleon mass at   normal density is
obtained. There are just two coefficients $x$ and $y$ used in the
scaling functions in SL1 and SL2. Without the inclusion of the
non-linear meson self-interacting terms, the saturation properties at
$\r_0=0.16$fm$^{-3}$ (see Table \ref{t:t1}) and the pressure within
the density region $\r=2-4.6\r_0$ are both nicely reproduced with the
SL1 and SL2 (see Fig.~\ref{pden}). It is worth mentioning that the
vector potential which is quadratic in density for the constant
$C_\om$ is known to result in a higher pressure above the
experimental region. The softening of the pressure here is attributed
to the contribution of rearrangement terms. For the SL1, only the
rearrangement term from the $\sg$ meson survives.

The parameter set SL3 does not consider the scaling of scalar meson
coupling constant but includes in the Lagrangian the non-linear $\sg$
self-interacting terms $U(\sg)=g_2\sg^3/3+g_3\sg^4/4$ with
coefficients $g_2=23.095$ and $g_3=-29.678$. Though these large
coefficients are needed to fit the saturation properties, they seem
to be inconsistent with the hypothesis of {\it
naturalness}~\cite{ma04,ue06}. In S3, we introduce a $g_2=-1.096$ to
obtain the given $\kappa$ in Table \ref{t:t1}. This is a little
different from the original one~\cite{song01}. As a comparison, we
can see in Fig.\ref{pden} that the SL3 and S3 parameter sets are not
consistent with the pressure constrained by the collective flow data.

\begin{figure}[tbh]
\begin{center}
\includegraphics[width=0.7\textwidth]{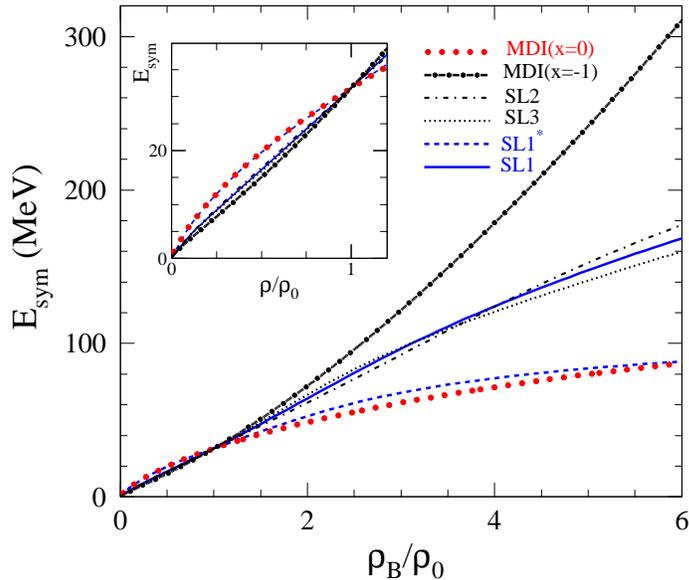}
 \end{center}
\caption{The symmetry energy as a function of density for different
models. The MDI(x=0) and MDI(x=-1) results are taken from
\cite{ch05}.\label{figse}}
\end{figure}

With (\ref{eqe1}), the symmetry energy in the RMF models can be
derived as
\begin{equation}\label{esym}
    E_{sym}=\frac{1}{2}\frac{\pp^2({\mathcal E}/\r)}{\pp\dt^2}
    =\frac{1}{2}C_\r^2\r +\frac{k_F^2}{6E_F}.
\end{equation}
It consists of contribution from the $\r$ meson (potential part)
and nucleons (kinetic part). In SL1 and SL2, we adopt the same
scaling function for coupling constants of $\r$ and $\om$ mesons:
$\Phi_\r(\r)=\Phi_\om(\r)=1-y\r/\r_0$. In this way, the ratio
$C_\r$ is just a constant that does not rely on the density. An
almost linear dependence  of the symmetry energy on the density is
expected from Eq.(\ref{esym}). The symmetry energy as a function
of density is shown in Fig.\ref{figse}. In Refs.~\cite{ch05,li05},
the symmetry energy extracted from the isospin diffusion data is
parameterized as $E_{sym}=31.6(\r/\r_0)^{\gm}$ with
$0.69\leq\gm\leq 1.05$ . All the parameter sets based on the BR
scaling discussed above lead to the symmetry energies between
those parameterized with $\gm=0.69$ and 1.05 in the whole density
region. In Ref.~\cite{li05}, the authors pointed out that with the
in-medium nucleon-nucleon cross sections, a symmetry energy of
$E_{sym}(\r) = 31.6(\r/\r_0)^{0.69}$ for $\r<1.2\r_0$ was found
most acceptable compared to the isospin diffusion data. However,
the symmetry energy at higher densities is not constrained at all.
It is thus interesting to examine predictions within the RMF
models with the chiral limit at high densities.

Our strategy is to first fit the symmetry energies constrained at low
densities by modifying the coefficient $y_\r$ in (\ref{sc3}), and
then predict the symmetry energy at high densities. The modified
symmetry energy is shown in Fig.\ref{figse} with the dashed blue
curve denoted as SL$1^*$. Considering the nucleon effective mass at
normal density is a little large in SL1, we introduce a coefficient
$y_\om$ in (\ref{sc3}) to reduce it to the value of $0.6M$. The new
parameter set is called SL$1^*$ and also listed in Table \ref{t:t1}.
The different parameterizations in SL1 and SL$1^*$ lead to
significantly difference in pressure at high densities as shown in
Fig.\ref{pden}. However, the suppression of the symmetry energy in
SL$1^*$ at high densities is dominated by the density dependence of
the ratio $C_\r$ in (\ref{esym}). This can be seen by comparing
results shown in Table \ref{t:t1} and Fig.\ref{figse} that the
density dependence of the symmetry energy does not change much by the
different nucleon effective masses of various parameter sets. In the
density-dependent RMF model~\cite{hof01}, the dropping of ratios like
the $C_\om$ or $C_\r$ is induced by the many-body correlation
contributions. Though the mechanism is different from the present
study, the correlation effect beyond the mean-field approximation may
play some role in obtaining the coefficients $y_\r$ and $y_\om$.

The symmetry energy at high densities from the MDI interaction with
x=0 is quite close to that given by the SL1$^*$. This is not
surprising since the SL1$^*$ is rendered to have the same symmetry
energies as the MDI(x=0) at low densities. On the other hand, this
justifies the consistency of the symmetry energy given by both models
at high densities. One can freely adjust $y_\r$ to fit the symmetry
energies given by the MDI(x=-1) model at low densities and then
examine the behavior at high densities. However, the $C_\r$ will
increase with density by doing so, which is contrary to the empirical
cases~\cite{hof01,fsu}. Therefore, based on the BR scaling the
symmetry energy at high densities is expected to be between those
given by the SL1 and the SL1$^*$.

\begin{figure}[tbh]
\begin{center}
\includegraphics[width=0.7\textwidth]{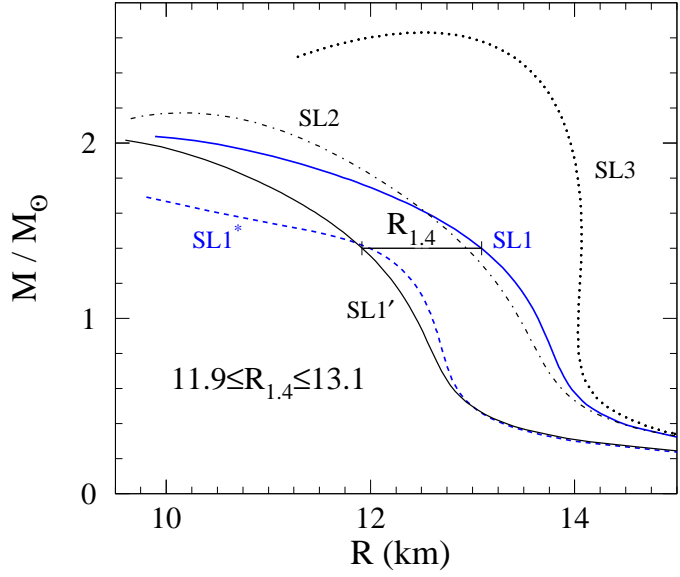}
 \end{center}
\caption{The neutron star mass versus the radius for various models.
The parameter set SL$1^\prime$ is just the same as the SL1 but with
$y_\r=0.526$. \label{fmsr}}
\end{figure}

We now turn to the astrophysical implications of the EOS's
constrained above. Generally speaking, these EOS's have the
special features of being soft at moderate densities and stiff at
high densities (see Fig.~\ref{pden}). It is thus interesting to
compare predictions using these EOS's with the recent
astrophysical observations. Recent studies on the millisecond
pulsar PSR J0751+1807 suggest that it has a mass of
$2.1\pm0.2(^{+0.4}_{-0.5})$ with $1\sg$ ($2\sg$)
confidence\cite{nic05}. Since the maximum mass of neutron star is
more sensitive to the EOS at high densities, this requires a
rather stiff EOS at least at high densities. In Fig.\ref{fmsr}, we
plot the mass-radius correlation of neutron stars obtained from
solving the standard TOV equation. In the models with chiral
limits, hadron masses approach zero at certain critical baryon
densities (see Table~\ref{t:t1}). The maximum central energy
density in a neutron star is that calculated at the maximum baryon
density. With SL1, the maximum neutron star mass is 2.04$M_\odot$
with a radius of R=9.89km. With SL2, these two observables are
2.17$M_\odot$ and R=10.13km, respectively. As the nucleon
effective mass at normal density is reduced by introducing the
parameter $y_\om$, the maximum neutron star mass also drops. This
is owing to that the introduction of $y_\om$ softens the EOS at
high densities and the energy density at the critical density is
thus lowered. With SL1$^*$, the nucleon effective mass is
$M_n^*=0.6M$ at $\r_0$, and the maximum neutron star mass is
$1.7M_\odot$. For a moderate value of $M_n^*=0.65M$, we can obtain
a moderately larger neutron star mass $1.94M_\odot$. Recent
measurements on the neutron star EXO 07482-676 gave a mass of
$M=2.10\pm0.28M_\odot$ and a radius of $13.8\pm1.8$km~\cite{oz06}.
The author indicated that the existence of such a massive neutron
star rules out all the soft EOS's of neutron-star matter. Among
the models studied here, only the parameter set SL3 can give
values close to the measurement for the EXO 07482-676. However,
the parameter set SL3 gives a pressure at $\r\leq4.6\r_0$ that is
too strong compared to that constrained by the collective flow
data in heavy-ion collisions~\cite{da02}. For the other parameter
sets, the radii of maximum mass neutron stars are just around
10km, which is below that given in \cite{oz06}, though the maximum
mass can be within the error bars of the measurement (except for
SL1$^*$).

Compared to the EOS obtained using the FSUGold parameter
set~\cite{fsu} the EOS's in the present study are stiffer at high
densities although they have the same incompressibility $\kappa=230$
MeV at normal density (see Fig.\ref{pden}). The EOS's obtained here
thus also result in larger maximum masses of neutron stars. In Ref.
\cite{li06}, the authors predicted a radius span of
11.5km$<$R$<$13.6km for the 1.4$M_\odot$ canonical neutron star.
Using the EOS's obtained in the present work, we obtain here a radius
span of 11.9km$<$R$<$13.1km for the 1.4$M_\odot$ neutron star, which
is comparable with that obtained in Ref. \cite{li06}.

\section{Summary}
In summary, within the RMF framework we have constructed several
new model Lagrangians using in-medium hadron properties according
to the Brown-Rho scaling due to the chiral symmetry restoration at
high densities. The model parameters are determined such that the
symmetric part of the resulting EOS's at supra-normal densities
are consistent with that required by the collective flow data from
high energy heavy-ion reactions, while the resulting density
dependence of the symmetry energy at sub-saturation densities
agrees with that extracted from the recent isospin diffusion data
from intermediate energy heavy-ion reactions. The rearrangement
terms are found to play an important role in softening the EOS at
moderate densities. The symmetry energy depends on the in-medium
hadron properties that are characteristic of the chiral symmetry
restoration at the critical density according to the BR scaling.
The resulting EOS's are then used to examine global properties of
neutron stars. It is found that the radius of a 1.4$M_\odot$
neutron star is in the range of 11.9km$\leq$R$\leq$13.1km.
Compared to other EOS's, the current EOS's have the special
feature of being soft at intermediate densities but stiff at high
densities naturally. This feature is important to produce a
heavier maximum neutron star mass around 2.0$M_\odot$ consistent
with recent observations.

\section*{Acknowledgement}
We thank P. Krastev and G.C. Yong for useful discussions. The work
was supported in part by the US National Science Foundation under
Grant No. PHY-0652548, the Research Corporation, the National Natural
Science Foundation of China under Grant Nos. 10405031, 10575071 and
10675082, MOE of China under project NCET-05-0392, Shanghai
Rising-Star Program under Grant No. 06QA14024, the SRF for ROCS, SEM
of China, and  the Knowledge Innovation Project of the Chinese
Academy of Sciences under Grant No. KJXC3-SYW-N2.

\end{document}